\documentclass[sigconf]{acmart}
\usepackage{array}
\usepackage{amsthm}
\usepackage{mathtools}
\usepackage{hyperref}
\usepackage{amsmath,amsfonts,bm}
\usepackage{subcaption,multirow,cleveref}
\crefname{algocf}{alg.}{algs.}
\Crefname{algocf}{Algorithm}{Algorithms}
\AtBeginDocument{%
  \providecommand\BibTeX{{%
    \normalfont B\kern-0.5em{\scshape i\kern-0.25em b}\kern-0.8em\TeX}}}

\AtBeginDocument{%
\providecommand\BibTeX{{%
 Bib\TeX}}}

\usepackage[ruled,linesnumbered]{algorithm2e}
\newcommand{\nonl}{\renewcommand{\nl}{\let\nl\oldnl}}

\begin{document}

% \title{Towards Decision Transformer based Offline Reinforcement Learning based Recommender Systems}
% \title{Diffusion-Based Policy Optimization with Energy Guidance for Offline Reinforcement Learning in Recommender Systems}
\title[Energy-Guided Diffusion Sampling in Reinforcement Learning-based Recommendation]{Energy-Guided Diffusion Sampling for Long-Term User Behavior Prediction in Reinforcement Learning-based Recommendation}
\author{Xiaocong Chen}
\affiliation{%
  \institution{Data 61, CSIRO}
  \city{Eveleigh}
  \country{Australia}
}
\email{xiaocong.chen@data61.csiro.au}
\author{Siyu Wang}
\affiliation{%
  \institution{Macquarie University}
  \city{Sydney}
  \country{Australia}}
\email{siyu.wang@mq.edu.au}

\author{Lina Yao}
\affiliation{%
  \institution{Data 61, CSIRO}
  \city{Eveleigh}
  \country{Australia}}
\affiliation{
  \institution{The University of New South Wales}
  \city{Sydney}
  \country{Australia}}
\email{lina.yao@data61.csiro.au}

\renewcommand{\shortauthors}{Xiaocong Chen, Siyu Wang, and Lina Yao}
\begin{abstract}
Reinforcement learning-based recommender systems (RL4RS) have gained attention for their ability to adapt to dynamic user preferences. However, these systems face challenges, particularly in offline settings, where data inefficiency and reliance on pre-collected trajectories limit their broader applicability. While offline reinforcement learning methods leverage extensive datasets to address these issues, they often struggle with noisy data and fail to capture long-term user preferences, resulting in suboptimal recommendation policies.
To overcome these limitations, we propose Diffusion-enhanced Actor-Critic for Offline RL4RS (DAC4Rec), a novel framework that integrates diffusion processes with reinforcement learning to model complex user preferences more effectively. DAC4Rec leverages the denoising capabilities of diffusion models to enhance the robustness of offline RL algorithms and incorporates a Q-value-guided policy optimization strategy to better handle suboptimal trajectories. Additionally, we introduce an energy-based sampling strategy to reduce randomness during recommendation generation, ensuring more targeted and reliable outcomes.
We validate the effectiveness of DAC4Rec through extensive experiments on six real-world offline datasets and in an online simulation environment, demonstrating its ability to optimize long-term user preferences. Furthermore, we show that the proposed diffusion policy can be seamlessly integrated into other commonly used RL algorithms in RL4RS, highlighting its versatility and wide applicability.
\end{abstract}


\maketitle

\section{Introduction}
Reinforcement Learning-based Recommender Systems (RL4RS) have become adaptive tools for addressing the complex challenges of personalized recommendations in domains such as e-commerce, advertising, and streaming~\cite{hu2018reinforcement,cai2018reinforcement,cai2017real,chen2022generative}. Unlike static algorithms, RL4RS dynamically learn by interacting with users and incorporating their feedback. However, their dependence on real-time interactions results in data inefficiency, as limited training interactions slow policy improvement and hinder adaptation to user behavior changes. To address this, offline reinforcement learning (offline RL) has emerged as a promising solution~\cite{Wang_2023,zhao2023user,chen2023opportunities,wang2025policy}. By leveraging pre-existing user interaction datasets, offline RL pre-trains RL4RS agents, reducing the reliance on live interactions and improving efficiency in environments with sparse or delayed feedback. For example, CDT4Rec~\cite{Wang_2023} integrates a Decision Transformer (DT) with causal mechanisms for reward estimation, while DT4Rec~\cite{zhao2023user} optimizes user retention through reward prompting tailored to RL4RS needs.

However, a critical limitation of previous offline RL-based methods in RL4RS is their reliance on imitating the behavior policy, which inherently biases the agent toward short-term interests~\cite{chen2023opportunities}. The behavior policy, which represents the strategy used to generate the offline dataset, often lacks the expressiveness needed to capture the full diversity of user preferences, particularly for long-term modeling. This limitation restricts the agent's ability to generalize beyond recent, short-term interactions emphasized in the dataset, making it difficult to identify and optimize for long-term trends in user behavior~\cite{zhu2023diffusion}. Given the dynamic nature of user preferences, which can change significantly over time, models that rely heavily on recent trajectories struggle to generalize to a broader spectrum of behavior, especially in rapidly evolving environments. As user interaction trajectories grow longer, these methods face increasing difficulties in extracting policies that perform well over extended horizons, ultimately leading to suboptimal recommendations that fail to sustain long-term user engagement. This highlights the need for more expressive behavior policies representation that can capture evolving trends and support long-term user satisfaction~\cite{chen2023deep}. In recommender systems, {\it expressiveness} refers to a policy's ability to model diverse and complex user behaviors, encompassing both immediate preferences and long-term trends.

\begin{figure}[h]
    \centering
    \includegraphics[width=0.8\linewidth]{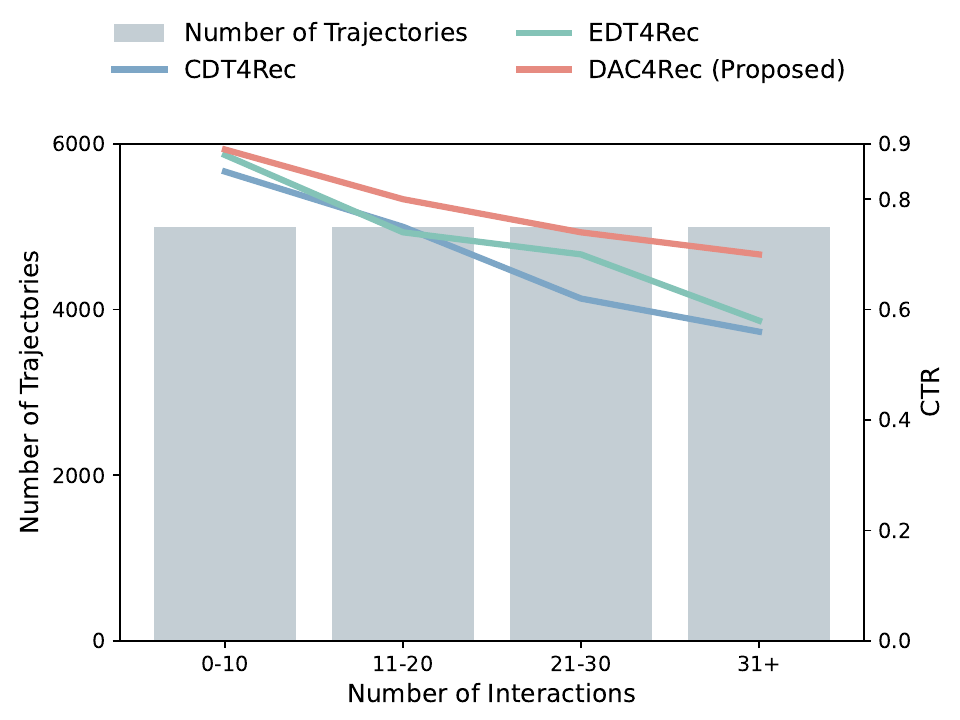}
    \caption{The illustration of the expressiveness challenges that were faced by offline RL4RS}
    \label{fig:traj}
\end{figure}

To illustrate the impact of behavior policy expressiveness on recommendation performance, we analyzed the performance of CDT4Rec~\cite{Wang_2023} and EDT4Rec~\cite{chen2024maximum} using the VirtualTB online simulation platform. As depicted in~\Cref{fig:traj}, the histogram reflects the distribution of user interaction trajectories across different groups, with each group containing an equal number of trajectories (5,000) to ensure a fair comparison. The line graph represents the corresponding recommendation performance, measured by Click-Through-Rate (CTR). We find that both CDT4Rec and EDT4Rec experience a significant decline in CTR as the length of user interaction histories increases. This result underscores the limitations of behavior policy expressiveness, as these models struggle to generalize to longer interaction trajectories. By failing to adequately capture the complexities of user behavior over extended histories, these methods produce suboptimal recommendations as interaction lengths increase.

Previous methods, such as those employed by~\citet{Wang_2023} and~\citet{chen2024maximum}, have relied on behavior cloning as a form of policy regularization to constrain the learned policy from deviating too far from the behavior policy. However, these approaches often restrict the exploration space to a narrow region, leading Q-learning to converge toward suboptimal policies~\cite{pearce2023imitating,he2024diffusion}. This limitation stems from the inability of traditional RL4RS methods to construct a strong and expressive representation of the behavior policy, which is crucial for capturing the complexities of user preferences.

Recent studies in diffusion-based imitation learning~\cite{pearce2023imitating} have demonstrated that diffusion models possess a strong capability to capture complex, multimodal features from offline datasets. Similarly, \citet{he2024diffusion} show that diffusion models effectively model intricate trajectories in offline datasets, highlighting their potential for improving behavior policy modeling in RL4RS. At the same time, diffusion models have gained significant attention in traditional recommendation systems, where they have been employed to construct robust representations and generate user preferences.

For example, diffusion models have been used to construct item representations in DiffuRec~\cite{li2023diffurec}, showcasing their superior representation capabilities. Additionally, \citet{liu2023diffusion} use diffusion to generate user preferences, demonstrating how diffusion models can effectively model and generate personalized recommendations. Despite these promising advancements in traditional recommendation systems, the use of diffusion models in RL4RS remains unexplored.

However, existing diffusion-based imitation learning approaches cannot be directly applied to the recommendation system domain. A major limitation is that these approaches assume the behavior policy is derived from expert demonstrations, implying that no better actions are available. This assumption does not hold true in recommendation systems, where user preferences are inherently sparse and dynamic. The recorded historical behaviors, which serve as expert demonstrations, are often influenced by intangible driving factors of human behaviors and may not represent optimal actions across different contexts or timeframes.

To address the limitations of existing offline RL4RS methods, we propose a novel approach that integrates diffusion models to enhance the expressiveness of behavior policy representations. We introduce a novel model named Diffusion-enhanced Actor-Critic for RL4RS (DAC4Rec) to address the challenges identified in offline RL4RS.

DAC4Rec builds upon the actor-critic structure and incorporates three key objectives: 1) a behavior-cloning term that encourages the model to sample actions from the same distribution as the training set, ensuring alignment with the historical behavior policy, 2) a policy improvement term that drives the model to sample high-value actions based on a learned Q-value, optimizing the recommendation process, and 3) An energy guidance mechanism is designed to reduce randomness in action selection, improving recommendations performance.  By conditioning the diffusion model on user states, DAC4Rec generates recommendations that not only closely align with the user's immediate preferences but also effectively model long-term preferences. This approach directly addresses the core challenges of RL4RS, particularly in dynamic environments where user preferences are constantly evolving.

Our contributions can be summarized as threefold:

\begin{itemize}
\item We introduce a novel model, Diffusion-Enhanced Actor-Critic for Reinforcement Learning-based Recommender Systems (DAC4Rec), to address the challenge of sub-optimal trajectories in offline RL4RS. Our approach leverages the diffusion process to effectively represent the behavior policy. Furthermore, we develop a robust energy-guided sampling strategy to mitigate randomness during the evaluation phase, ensuring more stable and reliable performance.

\item We validate the effectiveness of DAC4Rec through comprehensive experiments conducted on six publicly available datasets, as well as within an online simulation environment. The results consistently demonstrate its superior performance compared to existing methods. 
\item We demonstrate that the proposed diffusion policy can be easily integrated with a variety of widely used RL algorithms in RL4Rec, ensuring compatibility and enhancing their performance. 
\end{itemize}

\begin{figure*}[ht]
  \centering
  \includegraphics[width=0.85\linewidth]{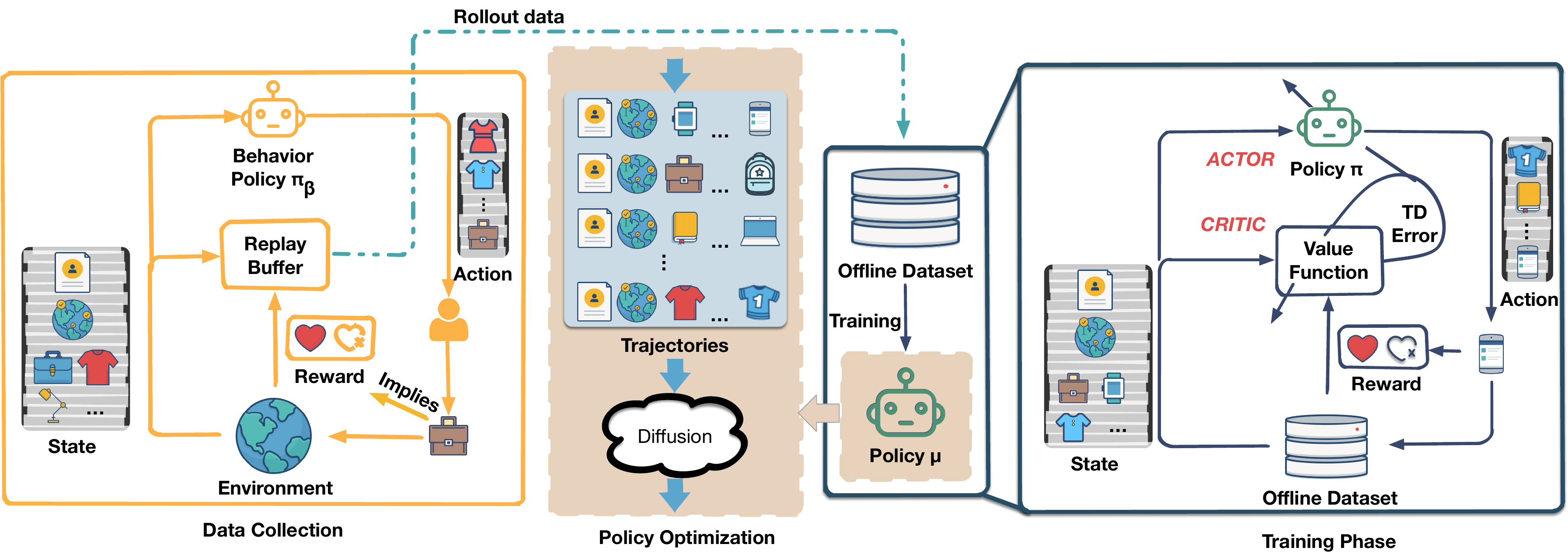}
  \caption{An overview of the offline RL4RS and the framework of the proposed DAC4Rec. The policy $\mu$ represents the proposed diffusion policy. While DAC4Rec is built on the actor-critic structure, it can be easily adapted to any Q-value-involved RL methods, including Q-learning, SAC, and DDPG (or TD3). }
  \label{fig:overrall}
\end{figure*}

\section{Background}
\subsection{Problem Formulation}
The recommendation problem can be viewed as an agent learning from user interactions (e.g., recommending items and receiving feedback) to achieve a goal. This fits naturally into the reinforcement learning (RL) framework, where the environment is modeled as a Markov Decision Process (MDP)~\cite{chen2023deep}, defined by the tuple $(\mathcal{S},\mathcal{A}, \mathcal{P}, \mathcal{R}, \gamma)$:

\begin{itemize}
    \item \textbf{State} $\mathcal{S}$: $s_t \in \mathcal{S}$ is the state at time $t$, containing user history or demographics.
    \item \textbf{Action} $\mathcal{A}$: $a_t \in \mathcal{A}(s_t)$ is the action in state $s_t$, typically recommended items.
    \item \textbf{Transition} $\mathcal{P}$: $p(s_{t+1}|s_t, a_t)$ is the probability of moving from $s_t$ to $s_{t+1}$ after $a_t$.
    \item \textbf{Reward} $\mathcal{R}$: $r(s, a)$ is the reward for taking action $a$ in state $s$.
    \item \textbf{Discount factor} $\gamma \in [0, 1]$: balances future and immediate rewards.
\end{itemize}

In \emph{offline} RL~\cite{levine2020offline}, the agent learns solely from a fixed dataset $\mathcal{D}$ without further interaction. For user $u$ at timestep $t$, $\mathcal{D}$ contains $(s_t^u, a_t^u, s_{t+1}^u, r_t^u)$, where $s_t^u$ is the user state, $a_t^u$ the recommended items, $s_{t+1}^u$ the next state, and $r_t^u$ the feedback.

\subsection{Diffusion Model}
\textbf{Forward Process.} 
Let $q(x)$ be a real data distribution, from which a sample $x_0 \sim q(x)$ is drawn. The forward diffusion process, defined as a fixed Markov chain, gradually introduces Gaussian noise to the sample over $T$ steps, resulting in a sequence of noisy samples $x_1, \dots, x_T$. The amount of noise added at each step is determined by a variance schedule $\beta_1, \dots, \beta_T$:
\begin{align} 
& q(x_t|x_{t-1}) = \mathcal{N}(x_t;\sqrt{1-\beta_t}x_{t-1},\beta_t \mathbf{I}) \nonumber\
& q(x_{1}|x_0) = \prod_{t=1}^T q(x_t|x_{t-1}). \label{eq:forward} 
\end{align} 
As $T \rightarrow \infty$, the distribution of $x_T$ converges to an isotropic Gaussian. Since every data sample is gradually turned into Gaussian noise, both the number of diffusion steps $T$ and the variance schedule $\{\beta_t\}_{t=1}^T$ can be adjusted to control the noise level at each step. However, learning the variance $\beta_t$ directly is difficult, so a reparameterization trick is employed to modify the forward process:
\begin{align} 
q(x_t|x_0)=\mathcal{N}(x_t;\sqrt{\overline{\alpha}_t}x_0,(1-\overline{\alpha}_t)\mathbf{I}), 
\end{align} 
where $\alpha_t = 1-\beta_t$ and $\overline{\alpha}_t=\prod_{s=1}^t \alpha_s$.

\textbf{Reverse Process.} 
Diffusion models learn a conditional distribution $p_\theta(x_{t-1}|x_t)$ and generate new samples by reversing the forward process: 
\begin{align} 
& p_\theta(x_{0})=p(x_T)\prod_{t=1}^T p_\theta(x_{t-1}|x_t), \nonumber \\ 
& p_\theta(x_{t-1}|x_t)=\mathcal{N}(x_{t-1};\mu_\theta(x_t,t),\Sigma_{\theta}(x_t,t)), 
\end{align} 
where $\mu_\theta(x_t,t)$ and $\Sigma_\theta(x_t,t)$ represent the mean and variance parameterized by $\theta$. For simplicity, $\Sigma_\theta(x_t,t)$ is set to $\sigma_t^2 \mathbf{I}$, where $\sigma_t^2=\frac{1-\overline{\alpha}_{t-1}}{1-\overline{\alpha}_t}$. The mean is given by: 
\begin{align} 
 \frac{1}{\sqrt{\alpha_t}} \bigg(x_t - \frac{\beta_t}{\sqrt{1-\overline{\alpha}_t}}\epsilon_\theta(x_t,t)\bigg). 
\end{align} 
In this way, the diffusion model can be primarily parameterized by $\epsilon\theta$.

Training is performed by maximizing the evidence lower bound (ELBO): 
\begin{align} 
\mathbb{E}_{x_0}[\log p_\theta(x^0)] \geq \mathbb{E}q \bigg[\log\frac{p_\theta(x^{0})}{q(x^{1
}|x^0)}\bigg]. 
\end{align} 
Once trained, sampling from the diffusion model involves drawing $x^T\sim p(x^T)$ and running the reverse diffusion chain to move from $t=T$ to $t=0$. Diffusion models can also be extended to conditional models by conditioning $p_\theta(x_{t-1}|x_t, c)$ on some condition $c$.

\section{Methodology} 

This section first details how the diffusion model is trained to approximate the behavior policy. Then, we discuss how the proposed method refines the learned policy to outperform the behavior policy by guiding the reverse diffusion process using both an energy function and a Q-value function. Specifically, DAC4Rec uses the energy function to guide the sampling toward high-reward actions while leveraging the Q-value function to perform policy improvement during the training of the behavior policy model. This section involves two distinct types of notations of steps when introducing the relevant method: i).\textbf{Diffusion step:} Associated with the forward and reverse diffusion process, denoted by superscript $i$, where $i \in \{1, \cdots, N\}$. These timesteps are used to model and refine the behavior policy as a generative process for user actions. ii). \textbf{Trajectory timesteps:} Related to the reinforcement learning sequence, indexed by subscript $t$, where $t \in \{1, \cdots, T\}$. These timesteps are used to optimize the target policy by evaluating and improving user interactions.

\subsection{Diffusion as a Policy} 
We define the behavior policy using the reverse process of a conditional diffusion model, parameterized by $\theta$, as follows:
\begin{align}
    \mathcal{N}(a^N;0,I)\sum_{i=1}^N p(a^{i-1}|a^i,s), \label{eq:piinbays}
\end{align}
where $p(a^{i-1}|a^i, s)$ follows a Gaussian distribution expressed as:
\[
\mathcal{N}(a^{i-1}; \sigma_\theta(a^i, s, i), \Sigma_\theta(a^i, s, i)),
\]
with $\Sigma_\theta(a^i, s, i) = \beta_i I$ representing the fixed covariance matrix. The mean is defined as:
\begin{align}
    \sigma = \frac{1}{\sqrt{\alpha_i}} \left(a^i - \frac{\beta_i}{\sqrt{1 - \overline{\alpha}_i}} \epsilon_\theta(a^i, s, i)\right).
\end{align}

To generate a sample, we initially draw $a^N$ from a standard normal distribution:
\[
a^N \sim \mathcal{N}(0, I).
\]
The reverse diffusion process, parameterized by $\theta$, is then applied iteratively as:
\begin{align}
   a^{i-1}|a^i = \frac{a^i}{\sqrt{\alpha_i}}-\frac{\beta_i}{\alpha_i\sqrt{1-\overline{\alpha}_i}}\epsilon_\theta(a^i,s,i) + \sqrt{\beta_i}\epsilon, \epsilon\sim\mathcal{N}(0,\mathbf{I}), \nonumber \\ \text{ for } i = N,\cdots,1. \label{eq:acondition}
\end{align}
 To improve the quality of the sampled actions, we set $\epsilon = 0$ when $i = 1$.

The objective function for training the conditional $\epsilon$-model is:
\begin{align}
    \mathcal{L}_d(\theta) = \mathbb{E}_{i \sim \mathcal{U}, \epsilon \sim \mathcal{N}(0, I), (s, a) \sim \mathcal{D}} \left[\|\epsilon - \epsilon_\theta(\sqrt{\overline{\alpha}_i} a + \sqrt{1 - \overline{\alpha}_i} \epsilon, s, i)\|^2\right], \label{eq:diffusionloss}
\end{align}
where $\mathcal{U}$ is a uniform distribution over $\{1, \cdots, N\}$.

This loss term, $\mathcal{L}_d(\theta)$, is designed to train the diffusion model as a policy that directly represents the behavior policy $\mu$, rather than merely imitating it. By leveraging the marginal distribution of the reverse diffusion process, this approach creates an expressive policy capable of capturing complex patterns, such as skewness and multi-modality, which are commonly present in offline datasets~\cite{zhu2023diffusion}. Furthermore, the sampling-based regularization only requires random samples from both $\mathcal{D}$ and the current policy, making it independent of the exact behavior policy—a significant advantage when handling datasets collected from human demonstrations.

Unlike conventional two-step approaches, this method provides a streamlined and effective way to regularize a flexible policy. The optimization of $\mathcal{L}_d(\theta)$ involves sampling from a single diffusion step $i$ for each data point. However, the reverse sampling process, requiring $N$ iterations of $\epsilon_\theta$, can slow down training. To address this, we may limit $N$ to a smaller value.

Using parameters $\beta_{\text{min}} = 0.1$ and $\beta_{\text{max}} = 10.0$, we define $\beta_i$ as suggested in \cite{song2020score}:
\[
\beta_i = 1 - \alpha_i = 1 - \exp\left(\beta_{\text{min}}\frac{1}{N} - 0.5(\beta_{\text{max}} - \beta_{\text{min}})\frac{2i - 1}{N^2}\right).
\]

\subsection{Diffusion Policy Training and Optimization} 
In this part, we focus on how the reverse diffusion process is utilized during training to learn and improve the behavior policy. To further refine this learned behavior policy, we integrate Q-learning into the training pipeline, enabling DAC4Rec to align the policy with high-reward actions.

While the policy regularization loss $\mathcal{L}_d(\theta)$ serves as a behavior-cloning term, it is insufficient on its own to enable the policy to outperform the behavior policy. To address this, we incorporate the Q-value function to guide the reverse diffusion process during the training of the behavior policy representation. The final policy learning objective is formulated as a weighted sum of both the policy regularization and improvement terms:

\begin{align} 
     \mu = \arg\min_{\mu_{\theta}}\mathcal{L}(\theta) = \mathcal{L}_d(\theta)+\mathcal{L}_q(\theta) \label{eq:overallpolicy}
\end{align} 
Let us now examine the policy improvement term. In a typical actor-critic framework for continuous control, Q-learning alternates between policy evaluation, which minimizes the temporal difference (TD) error $\mathcal{L}_{TD}(\phi)$, and policy improvement, which optimizes the policy: 
\begin{align} 
     \mathbb{E}_{s,a,s'\sim\mathcal{D}}[(r(s,a)+\gamma Q_{\phi'}(s',a')-Q_\phi(s,a))^2],
\end{align} 
where the next action is sampled from the current policy, $a'\sim\mu_\theta(\cdot|s')$. The goal is to maximize the expected Q values: \begin{align} 
    \max_\theta\mathbb{E}_{s\sim\mathcal{D},a\sim\mu_\theta(a|s)}[Q_\phi(s,a)].
\end{align}
While this objective is straightforward to optimize with a Gaussian policy, the sampling complexity of diffusion-based policies makes it more challenging. Therefore, we approach the offline RL problem through the lens of generative modeling, enabling diffusion policies to be trained in a supervised manner from the offline dataset. However, since offline RL datasets often contain suboptimal state-action pairs, the challenge lies in efficiently leveraging a well-trained Q-network $Q_\phi$ to guide the diffusion policy training.

By revisiting the forward diffusion process as in~\Cref{eq:forward}, we observe that the distribution of noisy actions $a^i$ at any step $i$ is given by: 
\begin{align} 
    q(a^i|a^0)=\mathcal{N}(a^i;\sqrt{\overline{\alpha}_i}a^0,(1-\overline{\alpha}_i)\mathbf{I}). 
\end{align} 
Using the reparameterization trick, we relate $a^i$, $a^0$, and $\epsilon$ as follows: 
\begin{align} 
    a^i=\sqrt{\overline{\alpha}_i}a^0 + \sqrt{1-\overline{\alpha}_i}\epsilon, \epsilon\sim\mathcal{N}(0,\mathbf{I}). 
\end{align} 
Since our diffusion policy predicts $\epsilon$ via $\epsilon_\theta(a^i,i;s)$, we approximate the action $a^0$ as: 
\begin{align} 
    \hat{a}^0 = \frac{1}{\sqrt{\overline{\alpha}_i}}a^i - \frac{\sqrt{1-\overline{\alpha}_i}}{\sqrt{\overline{\alpha}_i}}\epsilon_\theta(a^i,i;s).  \label{eq:a0}
\end{align} 

This approximation allows us to reconstruct $\hat{a}^0$ directly from $(s,a)$ which is a state-action pair that is sampled from the dataset, by first decomposing $a$ to $a^i$ and then applying a one-step denoising process. This approach, which we refer to as action approximation, simplifies the policy improvement step for diffusion policies: 
\begin{align} 
    \mathcal{L}_q(\theta)=-\mathbb{E}_{s\sim\mathcal{D},\hat{a}^0\sim\mu_\theta} [Q_\phi(s,\hat{a}^0)].  %L_q(\theta)
\end{align}

The policy improvement mechanism can be realized in practice through either direct policy optimization or likelihood-based policy optimization.

\textbf{Direct Policy Optimization.} In this method, Q-value maximization is achieved by backpropagating the gradients from the Q-network to the policy network: 
\begin{align} 
    \nabla_\theta \mathcal{L}_q(\theta) = -\frac{\partial Q_\phi(s,a)}{\partial a}\frac{\partial a}{\partial\theta}.
\end{align} 
This technique is feasible when $\frac{\partial a}{\partial\theta}$ can be computed. It is applicable to deterministic policies like DDPG and TD3, where the policy is parameterized.

\textbf{Likelihood-based Policy Optimization.} 
Alternatively, this approach transfers information from the Q-network to the policy indirectly by using:
\begin{align}
    \max_\theta \mathbb{E}_{s,a \sim \mathcal{D}}[g(Q_\phi(s,a))\log\mu_\theta(a|s)],
\end{align}
where $g(Q_\phi(s,a))$ is a function that increases with higher values of $Q_\phi(s,a)$, assigning more weight to actions with higher predicted values. In this work, we are using Advantage Weighted Regression~\cite{peng2019advantage} as the weight function.
This method requires the policy's log-likelihood to be both tractable and differentiable. For instance, IQL~\cite{kostrikov2021offline} follows this framework. It is a commonly used RL algorithm to address distributional shift and has potential benefits for the RL4RS community.

In contrast, the likelihood of samples in diffusion models is intractable. To handle this, we consider two alternative methods. First, instead of calculating the exact likelihood, we rely on a lower bound for $\log\mu_\theta(s,a)$, as suggested by the original DDPM~\cite{ho2020denoising}. By discarding constant terms that are independent of $\theta$, we obtain the following objective:

\begin{align} 
    \mathbb{E}_{i,\epsilon,(a,s)}\bigg[\frac{\beta_i\cdot g(Q_{\phi}(s,a))}{2\alpha_i(1-\overline{\alpha}_{i-1})}\|\epsilon-\epsilon_\theta(a^i,i;s)\|^2\bigg].
\end{align}

Second, we can substitute $\log\mu_\theta(s,a)$ with an approximated policy $\hat{\mu}_\theta(s,a) = \mathcal{N}(\hat{a}^0,\mathbf{I})$ based on~\Cref{eq:acondition}, leading to the following objective: 
\begin{align} 
    \mathbb{E}_{i,\epsilon,(a,s)}\Big[g(Q_\phi(s,a)\|a-\hat{a}^0\|^2)\Big].
\end{align}
Empirically, we observe that both approaches yield comparable performance; however, the latter is more straightforward to implement. Therefore, we primarily report results based on the second approach. 

By integrating this process into our diffusion policy optimization, the overall loss function for the policy improvement term $\mathcal{L}_q$ is defined as:
 
\begin{align}
    \mathbb{E}_{s_t,a_t,s_{t+1}\sim\mathcal{D},\hat{a}^0\sim\mu_{\theta}}[\|r(s_t,a_t)+\gamma Q_{\phi'}(s_{t+1},\hat{a}^0 )- Q_{\phi}(s_t,a_t)\|^2], \label{eq:qloss}
\end{align}
where $Q_{\phi_i'}$ denotes the target network.

\subsection{Guided Sampling}
Building on the reverse diffusion process used during training, we now focus on its role in sampling and evaluation. Unlike the training phase, where the reverse diffusion process is guided by optimization objectives, the sampling phase relies on randomly drawn actions from the learned policy distribution. 

Traditionally, continuous policies are modeled using a state-conditional Gaussian distribution. During evaluation, these policies typically operate deterministically by selecting the mean of the distribution as the action to minimize variance. In contrast, diffusion policies require sampling actions randomly from the underlying distribution, which lacks direct access to its statistics. This results in noisy sampling and high evaluation variance. A common solution involves the score function~\cite{song2020score}, but it depends on an additional classifier model, leading to significant computational overhead. To address this, we propose a novel approach tailored for RL4RS that maintains efficiency and high recommendation quality.

In offline RL, the objective is to learn a policy that maximizes cumulative returns or values. Even for stochastic policies like those in Soft Actor-Critic (SAC)~\cite{haarnoja2018soft}, the learned $Q_\phi$ can act as a deterministic evaluator for actions. By sampling multiple actions and selecting the one with the highest $Q_\phi$, we mitigate randomness. Actions are then chosen with probabilities proportional to $e^{Q(s,a)}$, effectively sampling from the diffusion policy $\mu(a|s)$ defined earlier.
\begin{align}
    \pi^*(a|s)\varpropto e^{Q(s,a)}\mu(a|s),
\end{align}
In order to sample actions from $\pi^*$ by diffusion sampling, we denote $\mu_0 := \mu$, $\pi_0 := \pi$, $a_0 := a$ at time $t = 0$\footnote{Since the diffusion process and the reinforcement learning process have been unified, $i$ and $t$ can be used interchangeably. For simplicity, we will use $t$ afterwards}. Then we construct a forward diffusion process to simultaneously diffuse $\mu_0$ and $\pi_0$ into the same noise distribution, where
\begin{align}
    \pi_{t0}(a_t|a_0, s) := \mu_t(a_t|a_0, s) = \mathcal{N}(a_t|\sqrt{\overline{\alpha}_t}a_0,(1-\overline{\alpha}_t)\mathbf{I}).
\end{align}
By doing so, we can have our action $a_t$ satisfy
\begin{align*}
    \pi_t(a_t|s)\varpropto e^{Q(s,a_t)}\mu_t(a_t|s), 
\end{align*}

The $e^{Q(s,a_t)}$ can be rewritten as an intermediate energy function $\mathcal{E}_t(s,a_t)$, 
\begin{align}
    & \mathcal{E}_t(s, a_t) = \log \mathbb{E}_{\mu_{0t}(a_0|a_t, s)} \left[ e^{Q_\phi(s, a_0)} \right] \\ \nonumber
    & \mathcal{E}_0(s, a_0) = Q_\phi(s, a_0). \nonumber
\end{align}

We now consider how to estimate the score function of $\pi_t(a_t|s)$ following the~\Cref{eq:overallpolicy}, 
\begin{align}
\nabla_{a_t} \log \pi_t(a_t|s) = \nabla_{a_t} \log \mu_t(a_t|s) + \underbrace{\nabla_{a_t} \mathcal{E}_t(s, a_t)}_{\approx f_\phi(s, a_t, t)}.    
\end{align}

To achieve the guided sampling process, $\nabla_{a_t} \mathcal{E}_t(s, a_t)$ serves as the desired guidance. Since this guidance depends on the $Q$-function which is already discussed in the previous section, we will need an extra neural network to estimate the targeted score function $\nabla_{a_t} \log \pi_t(a_t|s)$: An energy model $f_\phi(s, a_t, t)$ to estimate $\mathcal{E}_t(s, a_t, t)$ and guide the diffusion sampling process for $t > 0$ ($\mathcal{E}_t(s, a_t)$).

The energy guidance $\mathcal{E}_t(s,a_t)$ can be obtained by:
\begin{align}
    \mathbb{E}_{(s, a_0, t)} \left[ \left( f_\psi(s, a_t, t) - \log \mathbb{E}_{\mu_{0t}(a_0 | a_t, s)} \left[ e^{ Q_\phi(s, a_0)} \right] \right)^2 \right]. \label{eq:energy}
\end{align}

\subsection{Training Procedure} We implement the above methodology within an actor-critic framework, though it can be extended to other reinforcement learning algorithms, such as SAC and DDPG. The training pipeline for DAC4Rec is summarized in~\Cref{alg:overall_training}.
\begin{algorithm}[!ht]
\SetAlgoLined
\SetKwInOut{Input}{input}
\SetKwInOut{Output}{output}
\Input{Offline dataset $\mathcal{D}$, Behavior Policy Network $\mu_\theta$, critic network $Q_{\phi_1}$ and $Q_{\phi_2}$, target networks $\mu_{\theta'}$, $Q_{\phi_1'},Q_{\phi_2'}$, Energy Network $f_\phi$, and hyperparameters $\lambda$}

\tcp{Step 1: Train Diffusion-Based Behavior Policy}
\For{each iteration}{
Sample mini-batch $B = \{s_t, a_t, r_t, s_{t+1}\} \sim \mathcal{D}$ \;

\tcp{Q-Value Learning}
Sample $a_{t+1}^0 \sim \mu_{\theta'}(a_{t+1}|s_{t+1})$ by~\Cref{eq:acondition}\;
Update $Q_{\phi_1}$ and $Q_{\phi_2}$ using the~\Cref{eq:qloss}:

\tcp{Behavior Policy Learning}
Sample $a' \sim \mu_\theta(\cdot | s)$ \;
Calculate the action approximation by using~\Cref{eq:a0} \;
Update the policy network:
\[
\theta \leftarrow \theta - \eta \nabla_\theta \left( \mathcal{L}_d(\theta) + \lambda \mathcal{L}_q(\theta) \right),
\]
Update target networks $\theta', \phi_{1'}, \phi_{2'}$ \;
}

\tcp{Step 2: Train for Energy-Guided Sampling}
\For{each iteration}{
Sample mini-batch $B = \{s, a_0, t\} \sim \mathcal{D}$ \;
Define the energy function at $t=0$:

Train the energy model $f_\psi$ using the~\Cref{eq:energy}:

Update $f_\psi$ via gradient descent:

}

During reverse diffusion, guide the sampling process using the energy gradient:
\[
\nabla_{a_t} \log \mu_t(a_t | s)
\]
Generate sampled actions $a_0$ by iteratively applying reverse diffusion updates until convergence \;

\caption{Overall training algorithm for DAC4Rec}
\label{alg:overall_training}
\end{algorithm}

\section{Experiments}
In this section, we report the outcomes of experiments that focus on
the following five main research questions:
\begin{itemize}
    \item \textbf{RQ1}: How does DAC4Rec compare with existing offline RL-based recommendation system (RL4RS) methods and traditional deep RL algorithms in both online recommendation environments and offline datasets?
    \item\textbf{RQ2:} How does DAC4Rec perform in the situation when the user has longer interactions (i.e., the capability of capturing long-term preference) in the \emph{online} simulation environment?
    \item \textbf{RQ3:} What is the impact of hyper-parameter choices on DAC4Rec's performance in the \emph{online} simulation environment? 
    \item \textbf{RQ4:}How does each component of DAC4Rec contribute to the final performance in the \emph{online} simulation environment?
    \item \textbf{RQ5}: Can the proposed diffusion policy and energy-based sampling method be effectively integrated with other RL algorithms widely used in the RL4RS community?
\end{itemize}

\begin{table*}[!ht]
\centering
\caption{Performance comparison between selected RL methods trained with a user model on five datasets and online simulation environment. The best results are highlighted in bold. GRU is used as the state encoder for those offline datasets.}
\resizebox{0.96\textwidth}{!}{
\begin{tabular}{c|ccc|ccc|ccc}
\hline

\textbf{Method} & \multicolumn{3}{c|}{\textbf{Coat}} & \multicolumn{3}{c|}{\textbf{MoveLens}} & \multicolumn{3}{c}{\textbf{KuaiRec}} \\
 & $R_{cumu}$ & $R_{avg}$ & Length & $R_{cumu}$ & $R_{avg}$ & Length & $R_{cumu}$ & $R_{avg}$ & Length \\
\hline
\hline
DDPG & $16.3348 \pm 7.23$ & $2.3277 \pm 1.03$ & 7.02 & $9.3706 \pm 4.49$ & $3.0329 \pm 1.44$ & 3.11 & $9.2155 \pm 4.05$ & $1.0192 \pm 0.45$ & 9.04 \\
SAC & $80.2322 \pm 25.33 $&   $2.8321 \pm 0.89$  & 28.33 & $31.2491 \pm 8.77$&   $3.4642 \pm 0.97$  & 9.02 &  $23.1422 \pm 9.88$&  $0.7632 \pm 0.33$   & 30.32 \\
TD3 & $16.3232 \pm 7.02$ & $2.3542 \pm 1.01$ & 6.93 & $10.1620 \pm 4.90$ & $2.9410 \pm 1.42$ & 3.45 & $7.8179 \pm 3.25$ & $0.8610 \pm 0.36$ & 9.09\\
DT & $84.4224 \pm 26.82$ & $2.7916 \pm 0.89$ & 30.24 & $35.2451 \pm 9.80$ & $3.8135 \pm 1.06$ & 9.24 & $29.4353 \pm 10.23$ & $0.9070 \pm 0.32$ & 32.45 \\
DT4Rec & $82.3142 \pm 27.82$ & $2.6273 \pm 0.87$ & 31.33 & $33.6782 \pm 8.87$ & $3.2953 \pm 0.88$ & 10.22 & $27.8791 \pm 10.77$ & $0.8302 \pm 0.31$ & 33.58\\
CDT4Rec & $88.5823 \pm 23.53$ & $2.8698 \pm 0.76$ & 30.86 & $36.7256 \pm 10.24$ & $4.0935 \pm 1.14$ & 8.97 & $31.4452 \pm 11.42$ & $1.0075 \pm 0.37$ & 31.21 \\
EDT4Rec & $88.7854 \pm 22.88$ & $2.8622 \pm 0.74$ & 31.02 & $36.9821 \pm 9.92$ & $4.0551 \pm 1.09$ & 9.12 & $31.7281 \pm 10.88$ & $1.0478 \pm 0.36$ & 30.28\\
\hline
 \textbf{Ours} & $\mathbf{90.4222 \pm 23.53}$ & $\mathbf{2.8774 \pm 0.75}$ & 31.42 & $\mathbf{39.4217 \pm 11.02}$ & $\mathbf{4.2640 \pm 1.19}$ & 9.22 & $\mathbf{33.8210 \pm 12.02}$ & $\mathbf{1.0825 \pm 0.40}$ & 31.24\\
\hline
\end{tabular}
}

\resizebox{0.96\textwidth}{!}{
\begin{tabular}{c|ccc|ccc|ccc}

\hline

\textbf{Method} & \multicolumn{3}{c|}{\textbf{YahooR3}} & \multicolumn{3}{c|}{\textbf{KuaiRand}} & \multicolumn{3}{c}{\textbf{VirtualTB}} \\
 & $R_{cumu}$ & $R_{avg}$ & Length & $R_{cumu}$ & $R_{avg}$ & Length & $R_{cumu}$ & $R_{avg}$ & Length \\
\hline
DDPG & $7.7310 \pm 3.22$ & $2.675 \pm 1.11$ & 2.89 & $1.4232 \pm 0.51$ & $0.3287 \pm 0.12$ & 4.33 & $74.6534 \pm 24.22$ & $5.6470 \pm 1.83$ & 13.22 \\
SAC & $72.4372 \pm 21.44$&    $2.6612 \pm 0.79$   & 27.22 &  $6.5422 \pm 1.34$  &  $0.6011 \pm 0.60$    & 10.89 &  $72.3142 \pm 22.89$&   $5.2063 \pm 1.65$   & 13.89\\
TD3 & $8.2322 \pm 3.55$ & $2.8182 \pm 1.22$ & 2.92 & $1.5083 \pm 0.40$ & $0.3010 \pm 0.08$ & 5.01 & $75.2452 \pm 22.42$ & $6.0432 \pm 1.80$ & 12.45 \\
DT & $80.2746 \pm 20.47$ & $2.8688 \pm 0.73$ & 27.98 & $7.2653 \pm 2.01$ & $0.6749 \pm 0.18$ & 10.77 & $78.8762 \pm 23.99$ & $5.9706 \pm 1.83$ & 13.20 \\
DT4Rec & $78.9271 \pm 20.31$ & $2.7968 \pm 0.72$ & 28.22 & $6.8428 \pm 2.44$ & $0.6082 \pm 0.22$ & 11.25 & $76.6781 \pm 22.53$ & $5.5243 \pm 1.62 $ & 13.88\\
CDT4Rec & $81.2746 \pm 22.44$ & $3.3591 \pm 0.93$ & 24.21 & $7.4029 \pm 1.96$ & $0.6481 \pm 0.17$ & 11.42 & $79.7825 \pm 22.54$ & $5.6516 \pm 1.59$ & 14.12 \\
EDT4Rec & $82.4287 \pm 22.02$ & $3.4345 \pm 0.91$ & 24.00 & $7.6312 \pm 1.89$ & $0.6487 \pm 0.16$ & 11.77 & $79.9812 \pm 21.42$ & $5.7211 \pm 1.53$& 13.98\\
\hline
 \textbf{Ours} & $\mathbf{83.4627 \pm 21.87}$ & $\mathbf{3.5068 \pm 0.92}$ & 23.80 & $\mathbf{7.8802 \pm 2.07}$ & $\mathbf{0.6523 \pm 0.17}$ & 12.08 & $\mathbf{81.5261 \pm 21.29}$ & $\mathbf{6.1529 \pm 1.61}$ & 13.25 \\
\hline
\end{tabular}
}
\label{tab:allresult_1}
\end{table*}

\begin{table*}[!ht]
\centering
\caption{Performance comparison between selected RL methods trained with a user model on five datasets and online simulation environment. The best results are highlighted in bold. SASRec is used as the state encoder for those offline datasets.}
\resizebox{0.96\textwidth}{!}{
\begin{tabular}{c|ccc|ccc|ccc}
\hline

\textbf{Method} & \multicolumn{3}{c|}{\textbf{Coat}} & \multicolumn{3}{c|}{\textbf{MoveLens}} & \multicolumn{3}{c}{\textbf{KuaiRec}} \\
 & $R_{cumu}$ & $R_{avg}$ & Length & $R_{cumu}$ & $R_{avg}$ & Length & $R_{cumu}$ & $R_{avg}$ & Length \\
\hline
DDPG & $15.2764 \pm 7.02$ & $2.1186 \pm 0.97$ & 7.21 & $9.0271 \pm 4.32$ & $2.4721 \pm 1.18$ & 3.65 & $9.4018 \pm 4.24$ & $1.0197 \pm 0.46$ & 9.22 \\
SAC & $80.9823 \pm 24.22$ & $2.9008 \pm 0.78$ & 27.89 & $34.1192 \pm 8.92$ & $3.4624 \pm 1.12$ & 9.85 &  $23.4216 \pm 10.02$&  $0.7473 \pm 0.32$   & 31.34 \\
TD3 & $16.0280 \pm 7.48$ & $2.2832 \pm 1.06$ & 7.02 & $9.7262 \pm 4.66$ & $2.5791 \pm 1.24$ & 3.77 & $8.2458 \pm 3.89$ & $0.9154 \pm 0.43$ & 9.01 \\
DT & $83.4891 \pm 25.88$ & $2.7830 \pm 0.86$ & 30.02 & $33.2188 \pm 9.02$ & $3.4884 \pm 1.05$ & 9.52 & $29.0291 \pm 10.00$ & $0.9080 \pm 0.31$ & 31.98 \\
DT4Rec & $82.1823 \pm 26.22$ & $2.6156 \pm 0.83$ & 31.42 & $32.8728 \pm 8.91$ & $3.3238 \pm 0.91$ & 9.89 & $28.5418 \pm 9.51$ & $0.8798 \pm 0.29$ & 32.44\\
CDT4Rec & $87.8712 \pm 24.54$ & $2.9010 \pm 0.80$ & 30.29 & $35.1284 \pm 9.88$ & $4.0010 \pm 1.13$ & 8.78 & $30.4888 \pm 11.01$ & $1.0061 \pm 0.36$ & 30.22 \\ 
EDT4Rec & $88.1922 \pm 22.92$ & $2.8972 \pm 0.76$ & 30.42 & $35.8312 \pm 10.13$ & $3.9777 \pm 1.14$ & 8.89 & $31.0726 \pm 11.66$ & $1.0397 \pm 0.39$ & 29.88\\
\hline
\textbf{Ours} & $\mathbf{89.4721 \pm 22.48}$ & $\mathbf{2.9173 \pm 0.73}$ & 30.67 & $\mathbf{38.4718 \pm 10.72}$ & $\mathbf{4.2660 \pm 1.19}$ & 9.02 & $\mathbf{32.9899 \pm 11.87}$ & $\mathbf{1.0682 \pm 0.38}$ & 30.87 \\
\hline
\end{tabular}
}

\resizebox{0.96\textwidth}{!}{
\begin{tabular}{c|ccc|ccc|ccc}

\hline

\textbf{Method} & \multicolumn{3}{c|}{\textbf{YahooR3}} & \multicolumn{3}{c|}{\textbf{KuaiRand}} & \multicolumn{3}{c}{\textbf{VirtualTB}} \\
 & $R_{cumu}$ & $R_{avg}$ & Length & $R_{cumu}$ & $R_{avg}$ & Length & $R_{cumu}$ & $R_{avg}$ & Length \\
\hline
DDPG & $7.4421 \pm 3.12$ & $2.9063 \pm 1.22$ & 2.56 & $1.3972 \pm 0.52$ & $0.3288 \pm 0.12$ & 4.25 & $74.6534 \pm 24.22$ & $5.6474 \pm 1.83$ & 13.22 \\
SAC & $73.6721 \pm 21.01$&    $2.7748 \pm 0.79$   & 26.55 &  $6.8671 \pm 1.47$  &  $0.6590 \pm 0.14$    & 10.42 & $73.4271 \pm 21.92$&   $5.5125 \pm 1.65$   & 13.32\\
TD3 & $8.0623 \pm 3.35$ & $3.1243 \pm 1.30$ & 2.58 & $1.4765 \pm 0.47$ & $0.2828 \pm 0.09$ & 5.22 & $75.2452 \pm 22.42$ & $6.0442 \pm 1.80$ & 12.45 \\
DT & $79.8234 \pm 22.55$ & $2.8512 \pm 0.80$ & 28.04 & $6.9631 \pm 1.95$ & $0.6312 \pm 0.18$ & 11.03 & $78.8762 \pm 23.99$ & $5.9753 \pm 1.83$ & 13.20 \\
DT4Rec & $78.5172 \pm 23.02$ & $2.6688 \pm 0.78$ & 29.42 & $6.8172 \pm 1.82$ & $0.5686 \pm 0.15$ & 11.99 & $76.7871 \pm 22.47$ & $5.4420 \pm 1.59$ & 14.11\\
CDT4Rec & $80.7552 \pm 21.89$ & $3.2150 \pm 0.86$ & 25.41 & $7.3271 \pm 1.78$ & $0.6508 \pm 0.16$ & 11.27 & $79.2101 \pm 22.31$ & $5.6498 \pm 1.59$ & 14.02 \\ 
EDT4Rec & $82.4211 \pm 21.97$ & $3.4385 \pm 0.92$ & 23.97 & $7.5817 \pm 1.78$ & $0.6497 \pm 0.15$ & 11.67 & $79.6651 \pm 20.44$ & $5.6741 \pm 1.46$ & 14.04\\
\hline
\textbf{Ours} & $\mathbf{82.7564 \pm 23.66}$ & $\mathbf{3.4433 \pm 0.98}$ & 24.04 & $\mathbf{7.6672 \pm 1.86}$ & $\mathbf{0.6507 \pm 0.16}$ & 11.78 & $\mathbf{81.5261 \pm 21.29}$ & $\mathbf{6.1532 \pm 1.61}$ & 13.25 \\
\hline
\end{tabular}
}
\label{tab:allresult_2}
\end{table*}

We decide to validate the  RQ2-RQ5 on online simulation settings since they are more closely suited to real-world environments, whereas offline datasets are fixed and do not reflect users' dynamic interests.
\subsection{Datasets and Environments}
In this section, we evaluate the performance of our proposed algorithm, DAC4Rec, against other state-of-the-art algorithms, employing both real-world datasets and an online simulation environment. We first introduce five diverse, public real-world datasets from various recommendation domains for our offline experiments:
\begin{itemize}
    \item Coat~\cite{schnabel2016recommendations}, is used for product recommendation with 290 users, 300 items, and 7,000 interactions for training, and 290 users, 300 items, and 4,600 interactions for testing.

\item YahooR3~\cite{marlin2009collaborative}, is used for music recommendation with 15,400 users, 1,000 items, and 311,700 interactions for training, and 5,400 users and 54,000 interactions for testing.

\item  MovieLens-1M \footnote{https://grouplens.org/datasets/movielens/1m}: It contains 58,000 users, 3,000 items, and 5,900,000 interactions for training, and 58,000 users and 4,400,000 interactions for testing.

\item KuaiRec~\cite{gao2022kuairec}, is used for video recommendation with 1,000 users, 6,700 items, and 594,400 interactions for training, and 1,000 users and 371,000 interactions for testing.

\item KuaiRand~\cite{gao2022kuairand}, is similar to KuaiRec, but with 571,300 interactions for training, and 1,000 users and 371,000 interactions for testing.
\end{itemize}
All of those mentioned datasets are converted to the RL environment via EasyRL4Rec~\cite{yu2024easyrl4rec}. 

In addition, we also experiment on a real online simulation platform to validate the proposed method. We use VirtualTB~\cite{shi2019virtual} as the major online platform in this work. In terms of evaluation metrics, we use the same evaluation metrics as used on the offline datasets which are: $R_{cumu}, R_{avg}$ and Length. The $R_{cumu}$ represents the cumulative rewards received per trajectory and the $R_{avg}$ represents the average reward per step during this trajectory whose length is the length of the trajectory (i.e., number of interactions).

It is worth mentioning that, those offline datasets are converted to RL environments via the EasyRL4Rec framework which has a different setting than VirtualTB. Hence, the evaluation scale might be different.

\subsection{Baselines}
In our study, we focus on evaluating DAC4Rec within the context of offline RL4RS methods. Most existing works in RL4RS have been assessed using customized settings, which makes fair comparison challenging~\cite{chen2023deep}. To address this, we have selected a range of baselines, encompassing both prominent DT-based methods and well-established RL algorithms, for a comprehensive evaluation:
% Hence, we would like to provide two sets of 
\begin{itemize}
    \item \textbf{Deep Deterministic Policy Gradient (DDPG)}~\cite{lillicrap2015continuous}: An off-policy method designed for environments with continuous action spaces.
    
    \item \textbf{Soft Actor-Critic (SAC)}~\cite{haarnoja2018soft}: This approach is an off-policy, maximum entropy Deep RL method that focuses on optimizing a stochastic policy.

    \item \textbf{Twin Delayed DDPG (TD3)}~\cite{fujimoto2018addressing}: An enhancement of the baseline DDPG, TD3 improves performance by learning two Q-functions, updating the policy less frequently.

    \item \textbf{DT}~\cite{chen2021decision}: An offline RL algorithm that leverages the transformer architecture to infer actions.

    \item \textbf{DT4Rec}~\cite{zhao2023user}: Building on the standard DT framework, DT4Rec integrates a conservative learning method to better understand users' intentions in offline RL4RS.

    \item \textbf{CDT4Rec}~\cite{Wang_2023}: This model introduces a causal layer to the DT framework, aiming to more effectively capture user intentions and preferences in offline RL4RS.

    \item \textbf{EDT4Rec}~\cite{chen2024maximum}: This model is an extension of the CDT4Rec which considers the stitching problem in DT.
\end{itemize}

\begin{figure*}[!ht]
     \centering
     \begin{subfigure}[b]{0.3\linewidth}
         \centering
           \includegraphics[width=\linewidth]{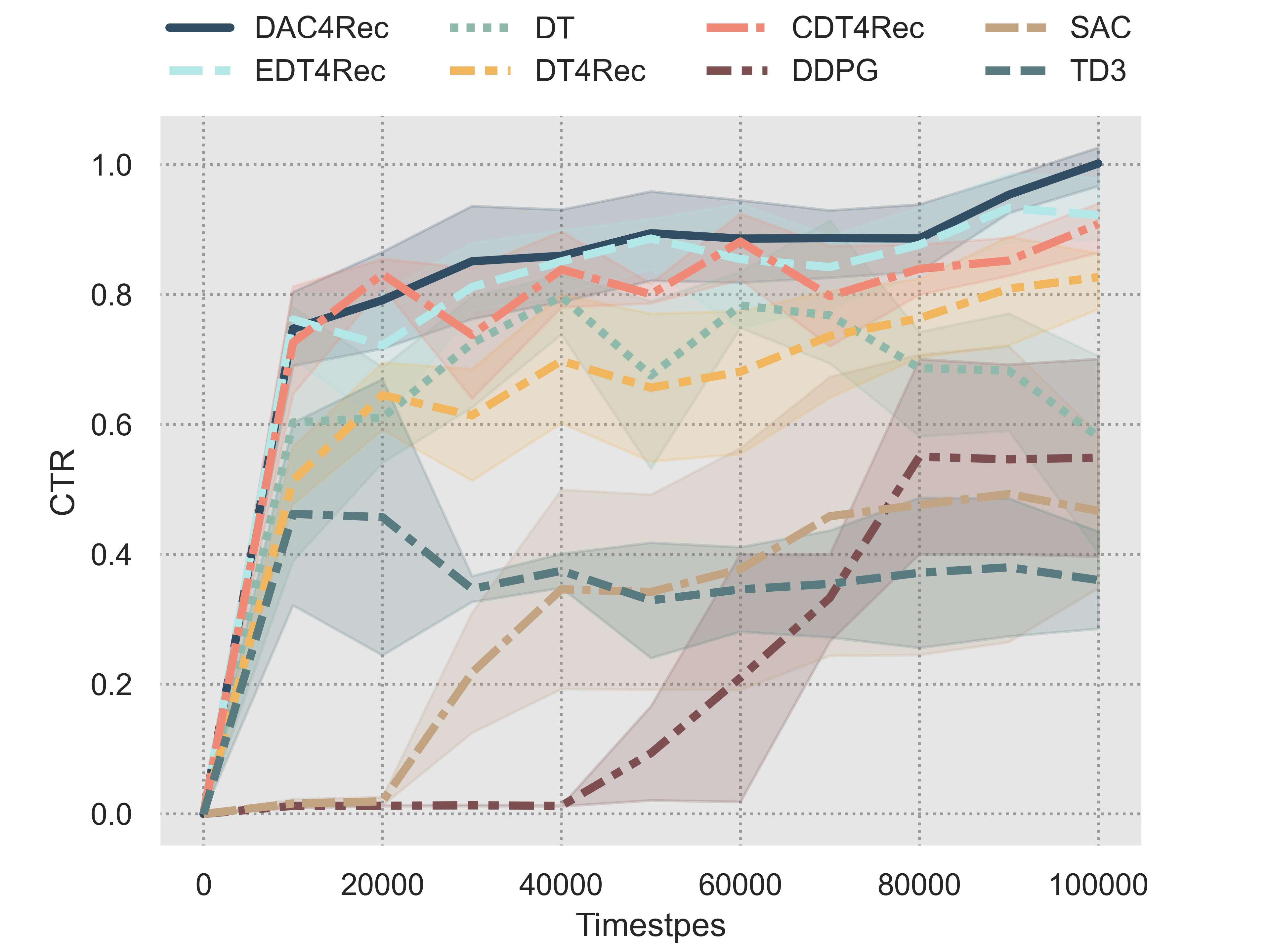}
           \caption{}
           \label{fig:over_comp}
     \end{subfigure}
     \begin{subfigure}[b]{0.3\linewidth}
         \centering
         \includegraphics[width=\linewidth]{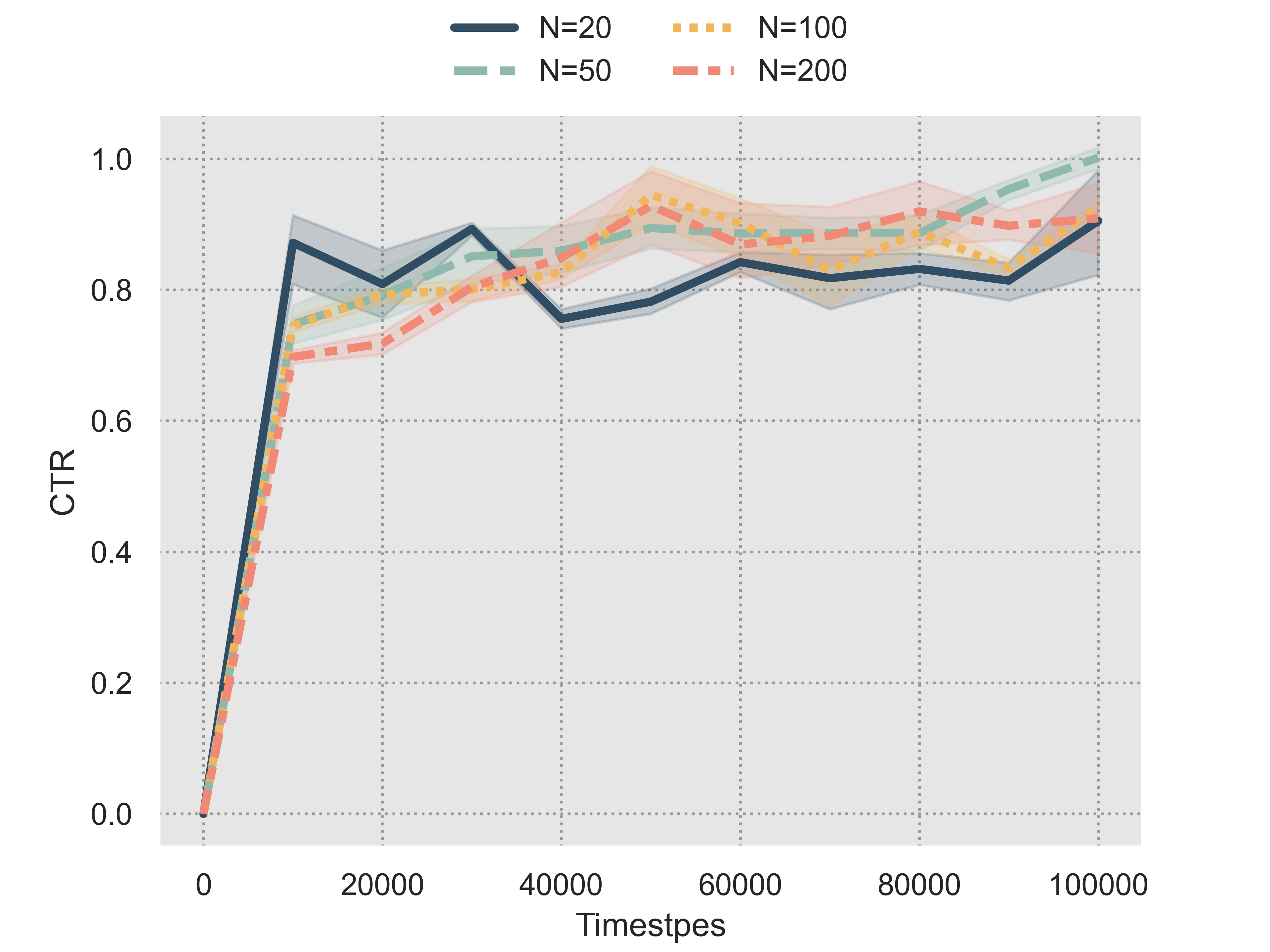}
         \caption{}
         \label{fig:different_n}
     \end{subfigure}
     \begin{subfigure}[b]{0.3\linewidth}
         \centering
         \includegraphics[width=\linewidth]{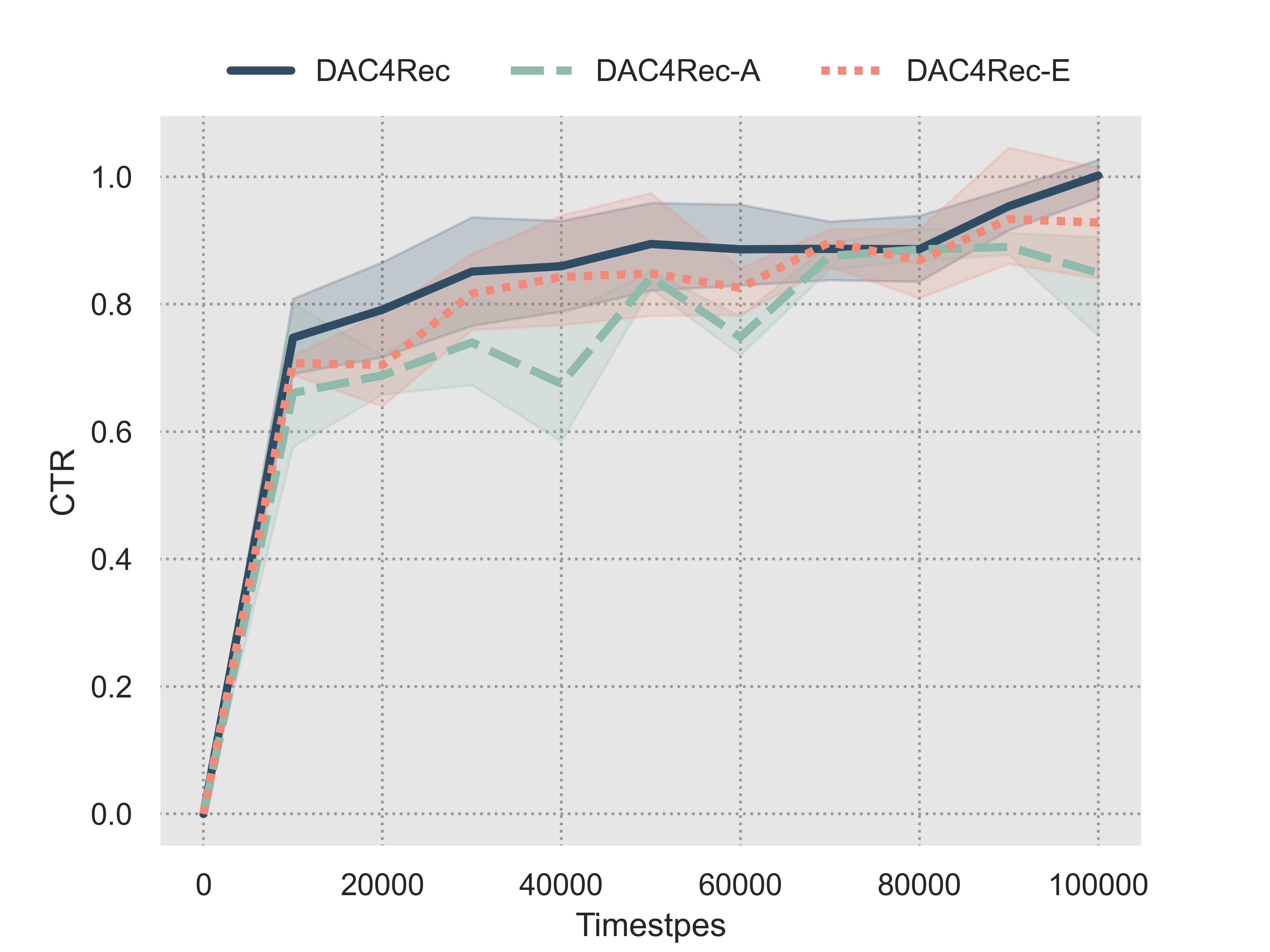}
         \caption{}
         \label{fig:ablation}
     \end{subfigure}
        \caption{(a). Overall comparison result with variance between the baselines and DAC4Rec in the VirtualTaobao simulation environment. (b).Hyperparameter diffusion step $N$ study, the value reported in the average CTR over $100,000$ timesteps. (c).Ablation study, the value reported in the average CTR over $100,000$ timesteps}
\end{figure*}

\subsection{Overall Comparison (RQ1)}
The summarized experimental results on the offline dataset are presented in~\Cref{tab:allresult_1,tab:allresult_2}, comparing RL methods across five datasets in offline and online simulation environments. Key metrics include cumulative reward (\(R_{\text{cumu}}\)), average reward per step (\(R_{\text{avg}}\)), and episode length. Among these, \(R_{\text{avg}}\) is the most informative as it normalizes rewards by the number of steps, providing a clearer measure of recommendation performance. In~\Cref{tab:allresult_1}, using GRU as the state encoder, the proposed method consistently achieves the highest \(R_{\text{cumu}}\) and \(R_{\text{avg}}\), highlighting its superior optimization of both cumulative and step-wise rewards. While SAC, DT4Rec, and EDT4Rec perform competitively on \(R_{\text{cumu}}\), they often lag in \(R_{\text{avg}}\), indicating less efficient step-wise optimization. \Cref{tab:allresult_2}, using SASRec as the state encoder, shows similar trends. The proposed method consistently outperforms others in \(R_{\text{avg}}\), demonstrating robustness across state encoders. Although DT4Rec and CDT4Rec slightly improve with SASRec, they remain behind the proposed method. For datasets like YahooR3 and VirtualTB, EDT4Rec improves step-wise performance but is still less effective than the proposed method.

\subsection{Long-Term Preference Study (RQ2)}
~\Cref{tab:length} presents a comparison of average CTR across three interaction lengths: 11–20, 21–30, and 31+ interactions, for four models—DT4Rec, CDT4Rec, EDT4Rec, and DAC4Rec. The results clearly indicate that DAC4Rec consistently outperforms the other models across all interaction lengths, demonstrating its superior ability to capture long-term user preferences.

DAC4Rec is particularly effective in maintaining stable performance as the interaction length increases, showing minimal decline in Avg. CTR even at longer interaction lengths. This stability suggests that DAC4Rec is well-suited for modeling users with stronger long-term preferences. In contrast, the other models experience a significant drop in performance as the interaction length grows, reflecting their limitations in adapting to extended user behaviors.

Moreover, DAC4Rec demonstrates lower variability in its performance compared to the other methods, further emphasizing its reliability and robustness. These results highlight DAC4Rec's strength in addressing the challenges of long-term preference modeling, especially in scenarios where users exhibit extended interactions and evolving preferences.

\begin{table}[!h]
\centering
\caption{Long-term Preference Experiments}
\begin{tabular}{c|ccc}
\hline
\multirow{2}{*}{Method} & \multicolumn{3}{c}{Avg.CTR}                                  \\ \cline{2-4} 
                           & \multicolumn{1}{c|}{11-20} & \multicolumn{1}{c|}{21-30} & 31+ \\ \hline
        DT4Rec          & \multicolumn{1}{c|}{$0.721 \pm 0.023$}      & \multicolumn{1}{c|}{$0.602 \pm 0.044$}      &  $0.545 \pm 0.028$   \\ \hline
        CDT4Rec         & \multicolumn{1}{c|}{$0.752 \pm 0.043$}      & \multicolumn{1}{c|}{$0.623 \pm 0.041$}      &  $0.564 \pm 0.039$    \\ \hline
        EDT4Rec         & \multicolumn{1}{c|}{$0.741 \pm 0.050$}      & \multicolumn{1}{c|}{$0.625 \pm 0.037$}      &  $0.587 \pm 0.042$  \\ \hline
        DAC4Rec         & \multicolumn{1}{c|}{$0.803 \pm 0.029$}      & \multicolumn{1}{c|}{$0.724 \pm 0.030$}      &  $0.702 \pm 0.022$   \\ \hline
\end{tabular}
\label{tab:length}
\end{table}
\subsection{Hyperparameter Study (RQ3)}
In this section, we analyze the hyperparameter $N$, which controls the number of diffusion steps in our proposed DAC4Rec model, to understand its impact on performance. Specifically, we perform an empirical study on how varying $N$ affects the model's performance on the online simulation platform VirtualTB. The results, illustrating the performance of DAC4Rec with different values of $N = \{20, 50, 100, 200\}$, are presented in~\Cref{fig:different_n}. 

The result shows that as $N$ increases, the performance becomes progressively more stable, with a noticeable reduction in variance. This suggests that higher values of $N$ allow the model to better approximate the target distribution, enhancing its robustness during the recommendation process. Among the tested values, $N=100$ achieves the best overall performance, striking an optimal balance between model accuracy and computational efficiency. While $N=200$ shows slightly higher stability, its gains in performance are marginal compared to $N=100$, which makes $N=100$ a more practical choice.

\subsection{Ablation Study (RQ4)}
In order to comprehensively understand the impact of action approximation and energy guidance on the proposed DAC4Rec model, we conducted an ablation study by evaluating two variations: DAC4Rec-E, which excludes energy guidance, and DAC4Rec-A, which omits action approximation. The results are presented in~\Cref{fig:ablation}. We observe that action approximation significantly contributes to the final performance by simplifying policy improvement and enabling more accurate action predictions. On the other hand, energy guidance shows limited improvement, aligning with its design purpose to enhance stochastic policies rather than deterministic ones like TD3. This finding reflects the compatibility of energy guidance with stochastic policy optimization, as deterministic policies naturally focus on single optimal trajectories. Further investigation into the role of energy guidance in stochastic policies is presented in the next section.

\subsection{Generalization on Various RL Backbones (RQ5)}
In this section, we investigate the generalization capability of the proposed diffusion policy and energy guidance algorithm when applied to widely used RL algorithms in recommender systems. Specifically, we focus on SAC and DDPG, which represent stochastic and deterministic policy-based RL algorithms, respectively. DT-based methods are excluded from this analysis as they do not rely on Q-value estimation. However, exploring their compatibility with diffusion-based approaches will be an exciting direction for future research, particularly in light of recent advancements~\cite{peebles2023scalable}.

The results are summarized in~\Cref{tab:different}. For DDPG, the addition of the diffusion policy significantly enhances average CTR compared to the baseline, while energy guidance alone provides a more modest improvement. The combination of diffusion and energy guidance achieves the best performance for DDPG, demonstrating their complementary effects. A similar pattern emerges for SAC, where diffusion yields substantial gains, and the integration of energy guidance further improves the results. Interestingly, energy guidance has a stronger positive impact on SAC than on DDPG, consistent with its design objective of reducing randomness and stabilizing stochastic policies. These findings illustrate the effectiveness of the proposed diffusion policy and energy guidance algorithm in generalizing across different RL backbones, particularly by enhancing performance and stabilizing policy behavior in stochastic settings.

\begin{table}[h]
    \centering
    \caption{Different Backbones Study}
    \begin{tabular}{c|c|c}
        \hline
         Method & Policy Type & Avg.CTR  \\ \hline
         DDPG & Deterministic & $0.7877 \pm 0.0625$\\
         DDPG+Diffusion& Deterministic & $0.8665 \pm 0.0453$\\
         DDPG+Energy & Deterministic& $0.7928 \pm 0.0538$ \\
         DDPG+Diffusion+Energy & Deterministic& $0.8771 \pm 0.0248$\\ 
         \hline
         SAC & Stochastic & $0.6824 \pm 0.0787$\\
         SAC+Diffusion & Stochastic&  $0.7927 \pm 0.0656$\\
         SAC+Energy & Stochastic& $0.7424 \pm 0.0431$\\
         SAC+Diffusion+Energy & Stochastic& $0.8455 \pm 0.0398$\\
         \hline
    \end{tabular}
    \label{tab:different}
\end{table}

\section{Related Work}  
\noindent\textbf{RL-based Recommender Systems.}
Notably,~\citet{zheng2018drn} pioneered the introduction of RL to enhance news recommender systems, and~\citet{zhao2018deep} further extended its application to the page-wise recommendation scenario.
Both of these approaches employ Deep Q-Networks (DQN)~\cite{mnih2013playing} to encode user and item information, effectively improving the quality of news recommendations. Moreover,~\citet{chen2020knowledge} incorporated knowledge graphs into the reinforcement learning framework, resulting in enhanced decision-making efficiency. Additionally,~\citet{chen2021generative} introduced a novel generative inverse reinforcement learning approach for online recommendations. This method autonomously extracts a reward function from user behavior, further advancing state-of-the-art in online recommendation systems.

\vspace{1mm}\noindent\textbf{Offline RL in RS.}
Recent studies have started exploring the prospect of integrating offline RL4RS~\cite{chen2023opportunities}. Notably,~\citet{Wang_2023,wang2024retentive} introduced a novel model called the Causal Decision Transformer for RS (CDT4Rec). This model incorporates a causal mechanism designed to estimate the reward function, offering promising insights into the offline RL4RS synergy. As an extension,~\citet{chen2024maximum} further explore the stitching capability of DT when dealing with the sub-optimal trajectories.Similarly,~\citet{zhao2023user} introduced the Decision Transformer for RS (DT4Rec), utilizing the vanilla Decision Transformer as its core architecture to provide recommendations, demonstrating its potential in the field.

\vspace{1mm}\noindent\textbf{Diffusion in RS}. Recent studies demonstrate that diffusion can significantly boost the recommendation performance. ~\citet{li2023diffurec} propose to use the diffusion model to construct the item representations in sequential recommendation. Differently,~\citet{liu2023diffusion} use the diffusion model to generate new user interactions to augment the data. Similarly,~\citet{ma2024plug} use the diffusion model to generate the user preferences to enhance the recommendation quality.
~\citet{qin2023diffusion} cause a diffusion-based sampling strategy to explore the user’s spatial visiting trends under the POI recommendation domain.
\section{Conclusion}
In this paper, we presented Diffusion-enhanced Actor-Critic for Offline RL4RS (DAC4Rec), a novel framework designed to address the limitations of traditional offline RL methods in recommender systems, particularly their reliance on suboptimal behavior policies. By integrating a denoising diffusion probabilistic model with reinforcement learning, DAC4Rec effectively models complex user preferences, enhancing robustness in both deterministic and stochastic policy settings. We also introduced a Q-value-guided policy optimization strategy and an energy-based sampling mechanism, which together improve recommendation quality and reduce randomness during generation.
Extensive experiments on six real-world offline datasets and in an online simulation environment demonstrated that DAC4Rec outperforms existing methods, particularly in scenarios with noisy or suboptimal trajectories. Furthermore, we showed that the proposed diffusion policy and energy-based sampling strategy are versatile and can be seamlessly integrated into other RL algorithms, such as SAC and DDPG, enhancing their performance in dynamic environments. Future work will explore the integration of diffusion policies into Decision Transformer-based methods to further expand the capabilities of RL4RS.

\section*{GenAI Usage Disclosure}
In accordance with the ACM authorship policy, no disclosure of generative AI usage is required for this work.
\bibliographystyle{ACM-Reference-Format}
\balance
\bibliography{sample-base}

\end{document}